\documentclass[conference]{IEEEtran}

\IEEEoverridecommandlockouts
\usepackage{cite}
\usepackage{amsmath,amssymb,amsfonts}
\usepackage{amsthm}
\usepackage{algorithmic}
\usepackage{graphicx}
\usepackage{textcomp}
\usepackage{xcolor}
\usepackage{hyperref}
\usepackage{makecell}
\usepackage{algorithm}
\usepackage{booktabs}
\usepackage{tabularx}
\usepackage{enumitem}
\usepackage{bbold}
\usepackage{algorithmic}
\usepackage{bbm}
\usepackage{quantikz} 
\usepackage{dblfloatfix} 

\newtheorem{definition}{Definition}

\newtheorem{proposition}{Proposition}
\newtheorem*{theorem*}{Theorem}

\hypersetup{
    colorlinks=true, 
    linkcolor=black,  
    citecolor=black, 
    filecolor=black, 
    urlcolor=black    
}

\def\BibTeX{{\rm B\kern-.05em{\sc i\kern-.025em b}\kern-.08em
    T\kern-.1667em\lower.7ex\hbox{E}\kern-.125emX}}
\begin{document}

\title{Consensus Protocols for Entanglement-Aware Scheduling in Distributed Quantum Neural Networks 
}

\author{
\IEEEauthorblockN{
    Kuan-Cheng Chen\IEEEauthorrefmark{2}\IEEEauthorrefmark{3}\IEEEauthorrefmark{1},
    Samuel Yen-Chi Chen \IEEEauthorrefmark{4},
    Mahdi Chehimi\IEEEauthorrefmark{5},
    Felix Burt \IEEEauthorrefmark{2}\IEEEauthorrefmark{3},
    Kin K. Leung\IEEEauthorrefmark{2}}

\IEEEauthorblockA{\IEEEauthorrefmark{2}Department of Electrical and Electronic Engineering, Imperial College London, London, UK}
\IEEEauthorblockA{\IEEEauthorrefmark{3}Centre for Quantum Engineering, Science and Technology (QuEST), Imperial College London, London, UK}
\IEEEauthorblockA{\IEEEauthorrefmark{4}Brookhaven National Laboratory, Upton, NY, USA}
\IEEEauthorblockA{\IEEEauthorrefmark{5}Department of Electrical and Computer Engineering, American University of Beirut, Beirut, Lebanon}
\IEEEauthorblockA{\IEEEauthorrefmark{1} Email: kuan-cheng.chen17@imperial.ac.uk}
}


\maketitle
\begin{abstract}
The realization of distributed quantum neural networks (DQNNs) over quantum internet infrastructures faces fundamental challenges arising from the fragile nature of entanglement and the demanding synchronization requirements of distributed learning. We introduce a \emph{Consensus–Entanglement-Aware Scheduling} (CEAS) framework that co-designs quantum consensus protocols with adaptive entanglement management to enable robust synchronous training across distributed quantum processors. CEAS integrates fidelity-weighted aggregation—where parameter updates are weighted by quantum Fisher information to suppress noisy contributions—with decoherence-aware entanglement scheduling that treats Bell pairs as perishable resources subject to exponential decay. The framework incorporates quantum-authenticated Byzantine fault tolerance, ensuring security against malicious nodes while maintaining compatibility with noisy intermediate-scale quantum (NISQ) constraints. Our theoretical analysis establishes convergence guarantees under heterogeneous noise conditions, while numerical simulations demonstrate that CEAS maintains 10–15 percentage points higher accuracy compared to entanglement-oblivious baselines under coordinated Byzantine attacks, achieving 90\% Bell-pair utilization despite coherence time limitations. This work provides a foundational architecture for scalable distributed quantum machine learning, bridging quantum networking, distributed optimization, and early fault-tolerant quantum computation.
\end{abstract}


\section{Introduction}
\label{sec:intro}

\begin{figure*}[!h]
  \centering
  \includegraphics[width=\textwidth]{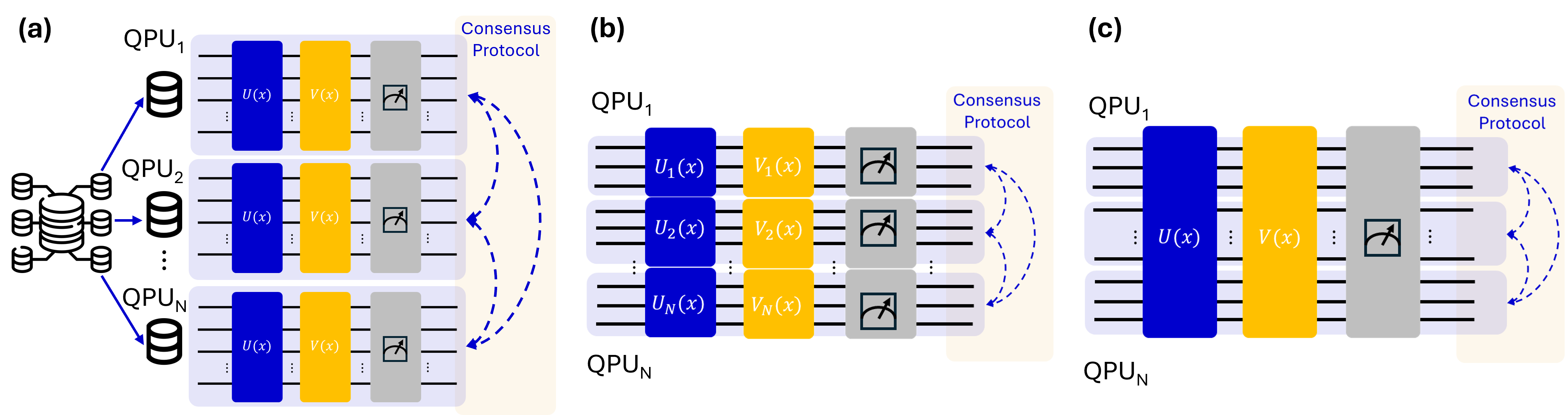}
  \caption{Consensus–entanglement–aware training topologies for three different DQNNs schemes.\textbf{(a)~Data-level partition.}  Each QPU hosts an \emph{identical} variational circuit $U(\boldsymbol{\theta})\,V(\boldsymbol{\theta})$ and measures local observables.  
A fidelity-weighted consensus layer (light yellow) aggregates the resulting gradients, distributing high-quality Bell pairs only to the most reliable links.  \textbf{(b)~Unitary-level partition.}  The global variational unitary is factorised into a sequential product
$U(\boldsymbol{\theta}) \;\approx\; U_{1}(\boldsymbol{\theta})\,U_{2}(\boldsymbol{\theta}) \,\cdots\, U_{N}(\boldsymbol{\theta})$.  
Each sub-unitary $U_{i}$ (and its local post-processing block $V_{i}$) is executed on a separate QPU; a layer-wise consensus round synchronises parameter updates before the forward state is teleported to the next node.  
\textbf{(c)~Circuit-level partition.}  A large qubit register is partitioned across QPUs.  Distributed entangling gates and non-local measurements are orchestrated through the consensus engine, enabling circuits whose width exceeds the native qubit count of any single device.  
}
  \label{fig:distqnn}
\end{figure*}

Recent advances in \emph{distributed quantum computing} have moved the vision of a global quantum infrastructure from concept to early-stage prototypes, unlocking the possibility of splitting large‐scale quantum workloads across spatially separated processors\cite{main2025distributed,cuomo2020towards,caleffi2024distributed,cacciapuoti2019quantum,barral2025review,van2016path}.  These processors must exchange quantum information with sufficiently high fidelity to preserve non-classical correlations, making the underlying network as critical to computational success as the devices themselves.  At the same time, \emph{quantum neural networks} (QNNs) have emerged as a promising algorithmic family capable of harnessing quantum parallelism for machine learning tasks that challenge classical resources\cite{abbas2021power,beer2020training,yu2024shedding,notton2025establishing}.  When QNNs are distributed across a quantum network\cite{hwang2025distributed,wu2024distributed,pira2023invitation}, their training phases rely on repeated exchange of quantum states and classical parameters\cite{chen2025quantum,liu2025federated,liu2025quantum,chen2025noise,chen2025federated,chen2025distributed,chen2025toward}, mandating rigorous coordination among nodes and efficient use of scarce entanglement.

Entanglement constitutes the foundational resource underpinning quantum networks, facilitating key primitives such as quantum teleportation, superdense coding, and non-local gate execution essential for distributed computation and gradient sharing\cite{cuomo2023optimized,ferrari2021compiler,wu2023qucomm,burt2024generalised,burt2025multilevel,burt2025entanglement,chen2026adaptive}. However, entanglement is both resource-intensive to generate and inherently susceptible to decoherence and operational noise, resulting in a critical trade-off between entanglement consumption and the fidelity constraints required to ensure reliable model convergence\cite{chen2025noise,ma2025robust}. As quantum networks grow and scale up in size, the orchestration of entanglement generation, routing, and utilization emerges as a primary performance bottleneck\cite{pant2019routing}, particularly in scenarios where multiple learning tasks compete for shared physical resources.

In classical distributed learning, global model consistency is secured through consensus protocols such as \emph{Byzantine fault tolerance} or \emph{synchronous averaging} schemes\cite{chen2025consensus}.  However, their direct quantum counterparts remain largely unexplored, even though the probabilistic nature of quantum operations, heterogeneous coherence times, and adversarial manipulation of quantum states exacerbate the very problems that consensus protocols aim to solve.  A \emph{quantum consensus protocol} must therefore harmonize classical message exchanges with quantum state manipulations while remaining robust to decoherence, gate errors, and malicious nodes capable of introducing stealthy phase rotations that cannot be detected by classical checksums alone\cite{chen2025noise}.

This paper presents a consensus-driven, entanglement-aware scheduling framework specifically developed to enable the efficient training of distributed QNNs, as shown in Fig.~\ref{fig:distqnn}. The proposed approach integrates fidelity-weighted consensus voting with dynamic entanglement brokerage, allowing each node to evaluate the quality of its outgoing quantum states and to allocate entanglement resources to training iterations where their informational value is highest. The framework characterizes the interplay between consensus latency and entanglement decoherence, demonstrating how coordinated scheduling can enhance convergence performance. Simulation results under realistic network noise conditions indicate substantial improvements in training efficiency compared to baseline entanglement-oblivious schedulers.

By articulating a cross-layer perspective that fuses quantum networking, machine learning principles, and distributed systems, this work lays foundational concepts for scalable, secure, and resource-efficient quantum learning across the emerging quantum Internet. Beyond immediate performance improvements, the proposed framework highlights a broader research agenda in which consensus mechanisms serve as the organizing logic for quantum network services, guiding the community toward rigorously engineered, experimentally validated deployments of large-scale distributed QNNs.

\section{Research Directions}

The effective distribution of quantum machine learning (QML) workloads across a network of quantum processing units (QPUs) presents fundamental challenges that necessitate integrated solutions across consensus, resource management, and security layers. This section outlines a research program centered on our Consensus--Entanglement-Aware Scheduling (CEAS) framework, which co-designs these layers to achieve robust, efficient, and scalable distributed QML.

\subsection{Integrated Control Plane for Distributed QML}

The practical realisation of distributed quantum learning confronts fundamental challenges rooted in quantum physics. Entanglement constitutes a perishable resource, as Bell pairs decohere exponentially, creating a critical race between consensus latency and quantum memory coherence times (\(T_1, T_2\)). This challenge is compounded by probabilistic gate and measurement errors that introduce heteroscedastic noise into gradient updates, potentially biasing the global model \cite{yuan2017fidelity}. Furthermore, hardware heterogeneity---spanning superconducting qubits, trapped ions, and photonic systems---precludes the application of uniform performance thresholds \cite{ladd2010quantum}, while covert noise and potential Byzantine faults manifest indistinguishably as systematic errors at the protocol level \cite{lo1999unconditional}. These factors collectively define a complex multi-objective optimization problem where latency, accuracy, and resource conservation are non-convexly coupled \cite{pirker2019quantum}.

Our proposed CEAS architecture addresses these challenges through a tightly integrated, cross-layer control plane. The consensus layer employs a hybrid gossip--Byzantine Fault Tolerance protocol, where fast gossip mixing is periodically checkpointed with a quorum of classically signed and quantum-authenticated states \cite{lo1999unconditional}. Lightweight fidelity witnesses, derived from local quantum Fisher information \cite{yuan2017fidelity}, act as weights during aggregation to suppress noisy contributions without full tomography. Complementarily, the scheduling layer treats entanglement as a perishable commodity, steering Bell pairs using regret-minimizing algorithms that balance routing latency against decoherence risk \cite{pirker2019quantum}. A bidirectional feedback interface enables the consensus engine to forecast entanglement demand while allowing the scheduler to signal authentication failures, facilitating real-time resource reallocation and adaptive trust management \cite{pirandola2020advances,chen2025noise}. This co-design ensures CEAS operates near the Pareto frontier of performance while respecting the stringent constraints of distributed quantum information processing.

\subsection{Robust Consensus with Fidelity Awareness}

A fundamental limitation of classical distributed learning in quantum contexts is its assumption of homogeneous update quality. In quantum networks, the fidelity of a node's gradient state---encoded in a quantum register---is degraded by variable coherence times, gate infidelities, and channel noise. To address this, we propose a consensus mechanism where aggregation is guided by quantitative fidelity measures \cite{yuan2017fidelity,khalid2024quantum}.

Each node computes a scalar fidelity stamp, \(\omega_k\), for its local parameter vector \(\boldsymbol{\theta}_k\). Suitable estimators include the Quantum Fisher Information (QFI), \(\mathcal{F}_k\), which bounds the precision of an unbiased estimator \cite{yuan2017fidelity}, or the process fidelity to an ideal channel \cite{khalid2024quantum}. The global model update is then computed as the fidelity-weighted mean:
\[
\bar{\boldsymbol{\theta}} = \frac{\sum_{k} \omega_k\,\boldsymbol{\theta}_{k}}{\sum_{k} \omega_k}, \quad \text{where } \omega_k \propto \phi(\mathcal{F}_{k}).
\]
This approach systematically suppresses the influence of low-fidelity, high-variance updates. A key research direction involves developing efficient, in-situ protocols for estimating \(\omega_k\) without resorting to full quantum state tomography \cite{khalid2024quantum}, thereby minimizing classical communication overhead.

\subsection{Decoherence-Aware Entanglement Scheduling}

The stochastic generation and finite coherence time of entanglement necessitate its treatment as a perishable, schedulable resource. Static allocation is suboptimal, leading to both resource starvation and decoherence-induced waste. We formulate the dynamic allocation of entanglement as a real-time scheduling problem \cite{pirker2019quantum,zhao2025qufm}.

The core objective is to map the demands of the consensus layer---dictated by the learning algorithm's step size---onto the physical layer's entanglement generation capabilities. This can be modeled as a Markov Decision Process (MDP) where the state space encompasses link fidelities, memory occupancy, and pending consensus deadlines. The action space involves prioritizing entanglement generation attempts across different network links. The reward function must balance multiple objectives: minimizing consensus latency, maximizing the fidelity of aggregated gradients, and conserving overall network resources. Investigating reinforcement learning agents, such as those based on Proximal Policy Optimization, to solve this MDP in simulated network environments such as NetSquid represents a promising path toward practical, adaptive schedulers \cite{pirker2019quantum}. Recent work on adaptive measurement policies \cite{zhao2025qufm} further enables dynamic trade-offs between estimation accuracy and measurement overhead as network conditions evolve.

\subsection{Byzantine Resilience with Quantum Authentication}

Distributed QML systems must be resilient to both benign faults and malicious (Byzantine) actors. Classical Byzantine Fault Tolerance (BFT) protocols are insufficient as they cannot directly verify the integrity of a quantum state. A quantum-native solution is required \cite{lo1999unconditional,pirandola2020advances}.

We propose integrating a quantum authentication step into the consensus protocol \cite{lo1999unconditional}. Each node tags its quantum gradient state \(\rho_k\) with an authentication key, typically in the form of entangled ancilla qubits. Any attempt to alter \(\rho_k\) will corrupt the authentication tag with high probability. The classical control plane can then verify the tag's syndrome. A state is accepted into the consensus quorum only if a supermajority of nodes (\(2f+1\) out of \(3f+1\)) validates the corresponding authentication syndrome and signs the associated classical metadata. This construction decouples safety---ensuring no corrupted quantum state is aggregated---from liveness. Open research problems include designing authentication schemes with low overhead, composably secure with the consensus logic, and analyzing their robustness under joint models of adversarial action and stochastic noise \cite{pirandola2020advances,chen2025noise}.

\subsection{Hardware-Aware Resource Management}

Practical quantum networks comprise heterogeneous hardware platforms with vastly different operational characteristics \cite{ladd2010quantum}. Nitrogen-vacancy centers offer millisecond coherence times but relatively slow gates, while superconducting qubits provide nanosecond gate speeds at the cost of microsecond-scale coherence. This heterogeneity necessitates adaptive resource management strategies that respect each platform's unique capabilities and limitations \cite{stein2023hetarch,wehner2018quantum}.

Our approach involves two key mechanisms: capability descriptors that provide coarse-grained, privacy-preserving summaries of each node's operational parameters \cite{stein2023hetarch}, and dynamic scheduling policies that adapt consensus cadence and entanglement routing based on these descriptors \cite{wehner2018quantum}. This hardware-aware control plane ensures latency-sensitive operations are directed toward fast, short-lived qubits, while long-coherence memories are reserved for tasks requiring extended quantum state storage. The challenge lies in developing efficient characterization protocols that provide sufficient information for optimal scheduling while preserving device privacy and minimizing characterization overhead.

\begin{figure*}[t!]
  \centering
  \includegraphics[width=0.85\textwidth]{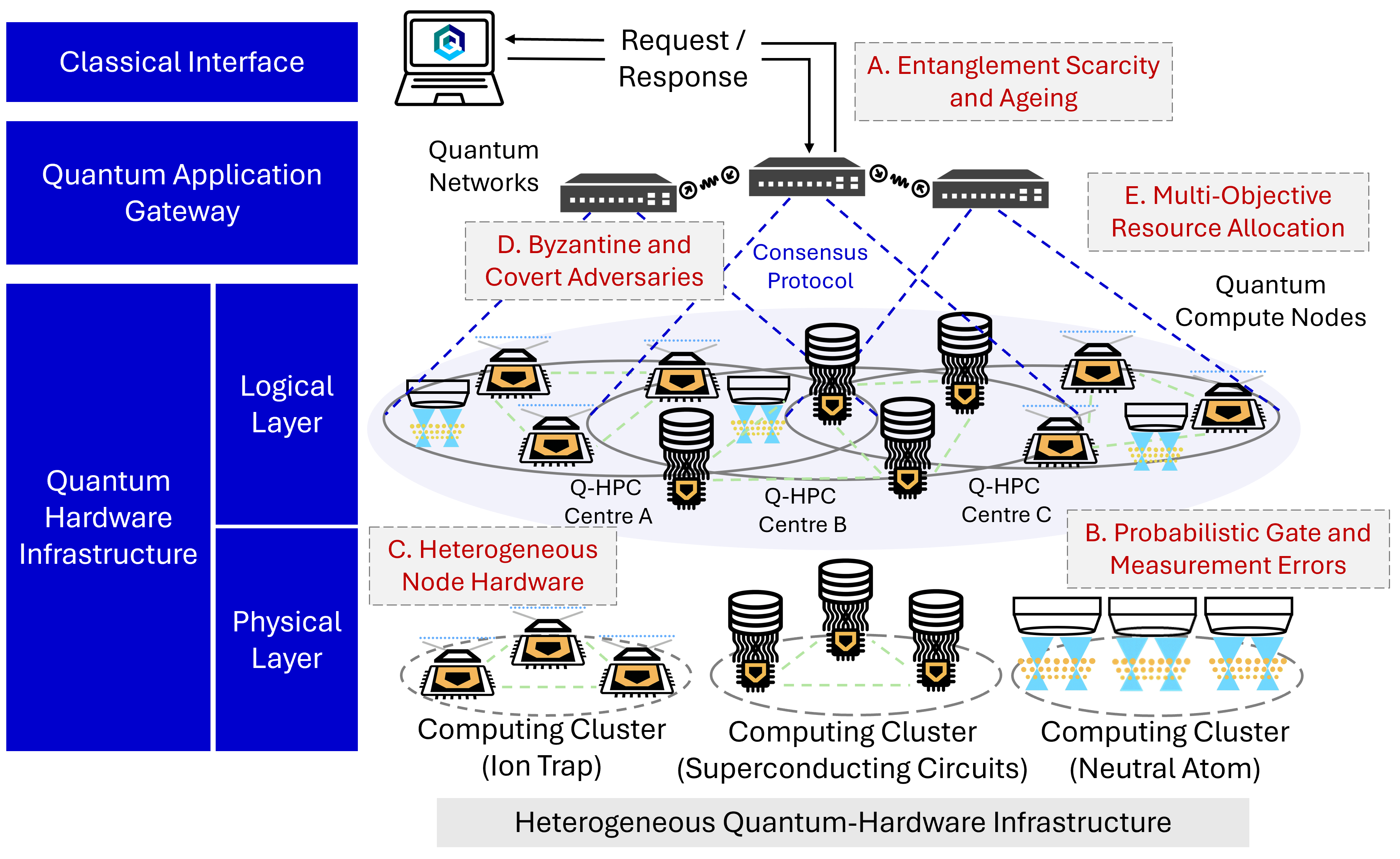}
  \caption{A co-designed architecture for distributed quantum neural network training. The logical layer executes a fidelity-aware consensus protocol, while the physical resource management layer handles entanglement generation and routing across heterogeneous hardware. A cross-layer broker mediates resource allocation based on learning demands.}
  \label{fig:quantum_network_architecture}
\end{figure*}

\section{Theoretical Framework}
\label{sec:theory}

This section formalises the CEAS framework. We begin with a system model for distributed quantum neural network (DQNN) training, then characterise fidelity-weighted consensus dynamics, entanglement-ageing and scheduling, and the adversarial model underpinning quantum Byzantine resilience. The goal is not to provide full finite-blocklength proofs, but to give a mathematically consistent abstraction that clarifies design choices and performance trade-offs.

\subsection{System Model}

We consider a set of $N$ quantum processing units (QPUs), indexed by $\mathcal{V} = \{1,\dots,N\}$, connected by a hybrid quantum--classical network whose topology at a given time is represented by a graph $G = (\mathcal{V},\mathcal{E})$. An edge $(i,j)\in\mathcal{E}$ denotes the ability to share Bell pairs between nodes $i$ and $j$ at a rate determined by underlying physical links and repeaters.

\subsubsection{Distributed Quantum Neural Network}

Let $\boldsymbol{\theta} \in \mathbb{R}^{d}$ denote the global parameter vector of a DQNN. For each node $k\in\mathcal{V}$ we associate a parametrised quantum channel
\[
\mathcal{E}_k(\boldsymbol{\theta}) : \rho_{\text{in}} \longmapsto \rho_{\text{out},k}(\boldsymbol{\theta}),
\]
implemented by a variational quantum circuit $U_k(\boldsymbol{\theta})$ acting on input state $\rho_{\text{in}}$ and followed by a measurement map $\mathcal{M}_k$. The local empirical loss at node $k$ is
\[
L_k(\boldsymbol{\theta}) \;=\; \mathbb{E}_{(x,y)\sim\mathcal{D}_k}\Bigl[\ell\bigl(y,\, h_k(x;\boldsymbol{\theta})\bigr)\Bigr],
\]
where $\mathcal{D}_k$ is the (possibly heterogeneous) data distribution, $\ell$ is a task-specific loss (e.g., cross-entropy), and $h_k$ denotes the classical post-processing applied to measurement outcomes.

The global objective is to minimise the aggregate loss
\begin{equation}
\label{eq:global_loss}
L(\boldsymbol{\theta}) \;=\; \frac{1}{N}\sum_{k=1}^{N} L_k(\boldsymbol{\theta}),
\end{equation}
which corresponds to the centralised training problem under ideal conditions. In CEAS, this optimisation is carried out collaboratively by nodes that exchange quantum states and classical summaries over $G$.

\subsubsection{Local Gradient Model}

In iteration $t$, node $k$ forms a stochastic estimate $\widehat{\boldsymbol{g}}_k^{(t)}$ of the global gradient $\nabla L(\boldsymbol{\theta}^{(t)})$ using variational techniques (e.g., parameter-shift rules) implemented on noisy quantum hardware. We use the decomposition
\begin{equation}
\label{eq:noise_model}
\widehat{\boldsymbol{g}}_k^{(t)} \;=\; \nabla L(\boldsymbol{\theta}^{(t)}) + \boldsymbol{\xi}_k^{(t)},
\end{equation}
where $\boldsymbol{\xi}_k^{(t)}$ captures the combined effects of (i) sampling noise, (ii) hardware noise, and (iii) channel noise in the distributed setting. We assume that for honest nodes $k\in\mathcal{H}$,
\[
\mathbb{E}\bigl[\boldsymbol{\xi}_k^{(t)} \mid \boldsymbol{\theta}^{(t)}\bigr] = \boldsymbol{b}_k^{(t)}, \qquad
\operatorname{Cov}\bigl[\boldsymbol{\xi}_k^{(t)} \mid \boldsymbol{\theta}^{(t)}\bigr] = \Sigma_k^{(t)},
\]
where the bias $\boldsymbol{b}_k^{(t)}$ and covariance $\Sigma_k^{(t)}$ depend on device and channel characteristics. For Byzantine nodes $k\in\mathcal{B}$ (defined below), $\widehat{\boldsymbol{g}}_k^{(t)}$ may be arbitrary.

A global parameter update at iteration $t$ takes the form
\begin{equation}
\label{eq:global_update}
\boldsymbol{\theta}^{(t+1)} \;=\; \boldsymbol{\theta}^{(t)} - \eta^{(t)}\,\widehat{\boldsymbol{g}}_{\text{agg}}^{(t)},
\end{equation}
where $\eta^{(t)}$ is the learning rate and $\widehat{\boldsymbol{g}}_{\text{agg}}^{(t)}$ is an aggregate of local estimates computed via CEAS.

\subsection{Fidelity-Weighted Consensus Dynamics}

The key theoretical idea in CEAS is that aggregation should depend on a reliability indicator for each node that reflects the quality of its quantum operations and channels.

\begin{definition}[Fidelity Stamp]
For each node $k$ and iteration $t$, a \emph{fidelity stamp} is a scalar statistic
\[
\phi_k^{(t)} \;=\; \phi\Bigl(\mathcal{F}_k^{(t)}, D_{\mathrm{PT},k}^{(t)}\Bigr) \in (0,1],
\]
constructed from:
\begin{itemize}[leftmargin=*]
    \item the quantum Fisher information $\mathcal{F}_k^{(t)}$ of the local circuit with respect to $\boldsymbol{\theta}^{(t)}$, and
    \item a process distance $D_{\mathrm{PT},k}^{(t)}$ between the implemented channel and the target ideal channel (e.g., a process-tensor or diamond-norm distance).
\end{itemize}
The function $\phi$ is chosen such that $\phi$ increases with $\mathcal{F}_k^{(t)}$ and decreases with $D_{\mathrm{PT},k}^{(t)}$.
\end{definition}

The fidelity stamp is computed locally and communicated as a low-dimensional classical quantity, typically with a few bits of precision. CEAS assigns a weight $w_k^{(t)}$ to node $k$ based on the stamp, for example
\begin{equation}
\label{eq:weight_def}
w_k^{(t)} \;\propto\; \phi_k^{(t)}, \qquad
\sum_{k\in\mathcal{V}} w_k^{(t)} = 1.
\end{equation}
The aggregate gradient is then
\begin{equation}
\label{eq:fidelity_weighted_grad}
\widehat{\boldsymbol{g}}_{\text{agg}}^{(t)} \;=\;
\sum_{k\in\mathcal{V}} w_k^{(t)}\,\widehat{\boldsymbol{g}}_k^{(t)}.
\end{equation}

The following proposition formalises, under simplifying assumptions, why fidelity weighting is desirable.

\begin{proposition}[Variance Reduction via Fidelity Weighting]
\label{prop:variance_reduction}
Suppose that for all honest nodes $k\in\mathcal{H}$ and iteration $t$,
\begin{enumerate}[label=(\roman*),leftmargin=*]
    \item $\mathbb{E}[\widehat{\boldsymbol{g}}_k^{(t)}] = \nabla L(\boldsymbol{\theta}^{(t)})$ (unbiasedness),
    \item $\operatorname{Cov}[\widehat{\boldsymbol{g}}_k^{(t)}] = \Sigma_k^{(t)}$ with $\Sigma_k^{(t)}$ positive definite, and
    \item gradient noise across honest nodes is uncorrelated.
\end{enumerate}
Consider linear aggregators of the form \eqref{eq:fidelity_weighted_grad} over honest nodes only. Among all choices of weights satisfying $\sum_{k\in\mathcal{H}}w_k^{(t)}=1$, the choice
\[
w_k^{(t)} \;\propto\; \bigl(\Sigma_k^{(t)}\bigr)^{-1}
\]
minimises the covariance of $\widehat{\boldsymbol{g}}_{\mathrm{agg}}^{(t)}$. In particular, if the scalar fidelity stamp is chosen such that $\phi_k^{(t)} \propto 1/\operatorname{tr}\bigl(\Sigma_k^{(t)}\bigr)$, then fidelity-weighted aggregation approximates the minimum-variance unbiased estimator.
\end{proposition}

\noindent
The proof follows from standard results on optimal linear combination of independent noisy estimators. In practice, $\Sigma_k^{(t)}$ is not directly accessible, but the quantum Fisher information $\mathcal{F}_k^{(t)}$ yields a lower bound on achievable variance; mapping $\phi_k^{(t)}$ monotonically to $\mathcal{F}_k^{(t)}$ therefore provides a principled heuristic for constructing near-optimal weights.

\subsubsection{Gossip-Based Mixing}

CEAS realises the aggregation in \eqref{eq:fidelity_weighted_grad} via a finite number of gossip exchanges over $G$. Let $\boldsymbol{\theta}_k^{(t,\ell)}$ denote the local parameter vector at node $k$ after $\ell$ gossip steps in iteration $t$. A generic update is
\begin{equation}
\label{eq:gossip_update}
\boldsymbol{\theta}_k^{(t,\ell+1)} \;=\;
\sum_{j\in\mathcal{N}(k)} W_{kj}^{(t,\ell)} \, \boldsymbol{\theta}_j^{(t,\ell)},
\end{equation}
where $\mathcal{N}(k)$ is the neighbourhood of $k$ (including $k$ itself) and $W^{(t,\ell)}$ is a row-stochastic mixing matrix whose entries incorporate fidelity weights and connectivity. Under mild connectivity and stochasticity conditions (e.g., $G$ is connected and $W^{(t,\ell)}$ is doubly stochastic),
\[
\boldsymbol{\theta}_k^{(t,\ell)} \;\xrightarrow[\ell \to \infty]{}\; \sum_{j\in\mathcal{V}} \pi_j^{(t)} \,\boldsymbol{\theta}_j^{(t,0)},
\]
where $\pi^{(t)}$ is the stationary distribution induced by the mixing process. By embedding fidelity stamps into $W^{(t,\ell)}$, CEAS ensures that nodes with persistently low fidelity have vanishing stationary weight, while nodes with reliable hardware dominate the consensus in the limit. The rate of convergence is governed by the spectral gap of $W^{(t,\ell)}$ and the connectivity of $G$.

\subsection{Entanglement-Ageing and Scheduling Model}

Entanglement is treated as a time-dependent resource subject to ageing and stochastic availability.

\begin{definition}[Entanglement Inventory]
For each edge $(i,j)\in\mathcal{E}$ and time $t$, let $M_{ij}(t)$ denote the number of Bell pairs stored in quantum memories associated with nodes $i$ and $j$. Each stored pair is annotated with its creation time $\tau$ and an estimated coherence time $\tau_c$, yielding a residual lifetime random variable
\[
T_{\mathrm{rem}}(t) \;=\; \max\{0,\,\tau + \tau_c - t\}.
\]
\end{definition}

A simple but analytically tractable model assumes that $T_{\mathrm{rem}}$ is exponentially distributed, with parameter $\lambda_{ij}$ determined by hardware and environmental conditions. The probability that a Bell pair remains usable over a consensus round of duration $\Delta t$ is then
\[
p_{ij}^{\mathrm{surv}}(\Delta t) \;=\; \mathbb{P}\bigl[T_{\mathrm{rem}} > \Delta t\bigr] \;=\; e^{-\lambda_{ij} \Delta t}.
\]

Let $r_{ij}(t)$ denote the rate at which the physical layer successfully generates new Bell pairs on link $(i,j)$ at time $t$, and let $c_{ij}^{(t)}$ denote the number of pairs consumed by the CEAS protocol in iteration $t$. Then the expected inventory dynamics satisfy
\begin{equation}
\label{eq:inventory_dynamics}
\mathbb{E}\bigl[M_{ij}(t+1)\bigr] \;=\;
e^{-\lambda_{ij}}\,\mathbb{E}\bigl[M_{ij}(t)\bigr] + r_{ij}(t) - c_{ij}^{(t)}.
\end{equation}
A feasible scheduling policy must ensure that the probability of a consensus failure due to lack of entanglement remains below an application-specific threshold.

\subsubsection{Scheduling as a Markov Decision Process}

Let $s_t$ be a state vector containing, for all $(i,j)\in\mathcal{E}$, the current estimated inventories $M_{ij}(t)$, decay parameters $\lambda_{ij}$, and relevant performance indicators (e.g., backlog of pending consensus rounds). A CEAS scheduling policy chooses an action $a_t$ specifying:
\begin{itemize}[leftmargin=*]
    \item generation rates $r_{ij}(t)$ subject to hardware constraints, and
    \item allocation decisions $c_{ij}^{(t)}$ for each consensus request.
\end{itemize}
The transition $s_t \mapsto s_{t+1}$ is stochastic, driven by the randomness of entanglement generation and decoherence, but follows probabilistic dynamics consistent with \eqref{eq:inventory_dynamics}. A reward function $R(s_t,a_t)$ captures the trade-off between training progress (e.g., reduction in global loss $L$ or improvement in test accuracy) and resource cost (e.g., expected number of Bell-pair generation attempts).

\begin{definition}[CEAS Scheduling MDP]
The entanglement scheduling problem is the Markov decision process (MDP) defined by the tuple $(\mathcal{S},\mathcal{A},P,R,\gamma)$, where:
\begin{itemize}[leftmargin=*]
    \item $\mathcal{S}$ is the state space of inventories and performance indicators;
    \item $\mathcal{A}$ is the space of generation and allocation decisions;
    \item $P(s_{t+1} \mid s_t,a_t)$ is induced by entanglement generation and decoherence processes;
    \item $R(s_t,a_t)$ encodes the time--accuracy--cost trade-off; and
    \item $\gamma\in(0,1)$ is a discount factor capturing the importance of future performance.
\end{itemize}
\end{definition}

In this abstraction, CEAS seeks a policy $\pi^*$ that maximises the expected discounted return $\mathbb{E}\bigl[\sum_{t=0}^{\infty} \gamma^t R(s_t,a_t)\bigr]$. In practice, the policy is obtained via approximate dynamic programming or deep reinforcement learning, and is constrained to operate on coarse-grained state features to keep overhead manageable.

\subsection{Adversarial Model and Quantum Byzantine Resilience}

To capture malicious or misconfigured behaviour, we partition the node set into honest and Byzantine subsets, $\mathcal{V} = \mathcal{H} \cup \mathcal{B}$, with $\mathcal{H}\cap\mathcal{B}=\emptyset$.

\begin{definition}[Byzantine Adversary]
A node $k\in\mathcal{B}$ is called Byzantine if, at any iteration $t$, it may produce arbitrary classical messages and quantum states, possibly in a coordinated manner with other Byzantine nodes and with knowledge of the protocol state. The adversary is computationally unbounded but cannot break the laws of quantum mechanics.
\end{definition}

We adopt the standard assumption that the number of Byzantine nodes satisfies $|\mathcal{B}| \leq f < \lfloor N/3\rfloor$, which is necessary for deterministic consensus with authenticated channels in the classical setting. Quantum-specific corruption modes include:
\begin{itemize}[leftmargin=*]
    \item injecting malicious gradient states that deviate significantly from the true optimisation dynamics;
    \item performing measurements or entanglement-breaking operations on shared Bell pairs to degrade fidelity; and
    \item replay or forging of classical metadata associated with quantum payloads.
\end{itemize}

\subsubsection{Authentication Tags and Trust Scores}

CEAS couples classical BFT logic with quantum authentication to ensure that only high-integrity updates influence the global model. Let $\rho_k^{(t)}$ denote the quantum payload from node $k$ at iteration $t$ (e.g., a gradient-encoding state or a teleported intermediate feature state). An authentication tag is an ancillary system $\sigma_k^{(t)}$ entangled with $\rho_k^{(t)}$ via an encoding map $\mathcal{A}_k$, such that any tampering with $\rho_k^{(t)}$ induces a detectable change in the joint state.

Upon reception, a verification map $\mathcal{V}_k$ produces a binary outcome $v_k^{(t)}\in\{0,1\}$ indicating authentication success or failure. The probability of a false negative (accepting a tampered state) can be made exponentially small in the length of the tag, at the cost of additional qubits and operations.

CEAS maintains a trust score $\tau_k^{(t)}\in[0,1]$ for each node, updated according to a rule such as
\[
\tau_k^{(t+1)} \;=\; \alpha\,\tau_k^{(t)} + (1-\alpha)\,v_k^{(t)},
\]
where $\alpha\in(0,1)$ controls the memory of past behaviour. Nodes whose trust score falls below a threshold $\tau_{\min}$ are quarantined: their contributions are assigned weight $w_k^{(t)}\approx 0$ in \eqref{eq:fidelity_weighted_grad}, and they may be excluded from BFT quorums. Error-mitigation and diagnostics routines can be used locally to attempt recovery; once the node’s fidelity and trust indicators improve, it can gradually re-enter the consensus process.

\subsubsection{Safety and Liveness Guarantees}

At a high level, CEAS seeks to preserve two classical properties:
\begin{itemize}[leftmargin=*]
    \item \emph{Safety}: no global update is committed that depends on undetected, maliciously altered quantum payloads or forged classical metadata.
    \item \emph{Liveness}: honest nodes eventually make progress in training, provided the network remains sufficiently connected and entanglement resources remain above a minimal level.
\end{itemize}

Under standard assumptions on authenticated classical channels, bounded $f<\lfloor N/3\rfloor$, and sufficiently strong quantum authentication codes, the combination of BFT quorums and trust-score-based quarantine yields safety guarantees analogous to those in classical Byzantine consensus, with additional robustness against covert quantum tampering. Liveness is ensured by design of the BFT checkpoint cadence and by entanglement scheduling policies that avoid starving honest nodes. A full composable security analysis in the universal composability framework is left as future work, but the abstractions above provide a consistent foundation for protocol design.

\medskip
Overall, the CEAS theoretical model links physical-layer parameters (error rates, coherence times, entanglement generation rates) to consensus-layer quantities (fidelity weights, mixing matrices, trust scores) and, ultimately, to optimisation behaviour (variance of aggregated gradients, convergence rates). This cross-layer linkage underpins the architectural decisions discussed in subsequent sections.

\begin{figure}[!h]
  \centering
  \includegraphics[width=0.5\textwidth]{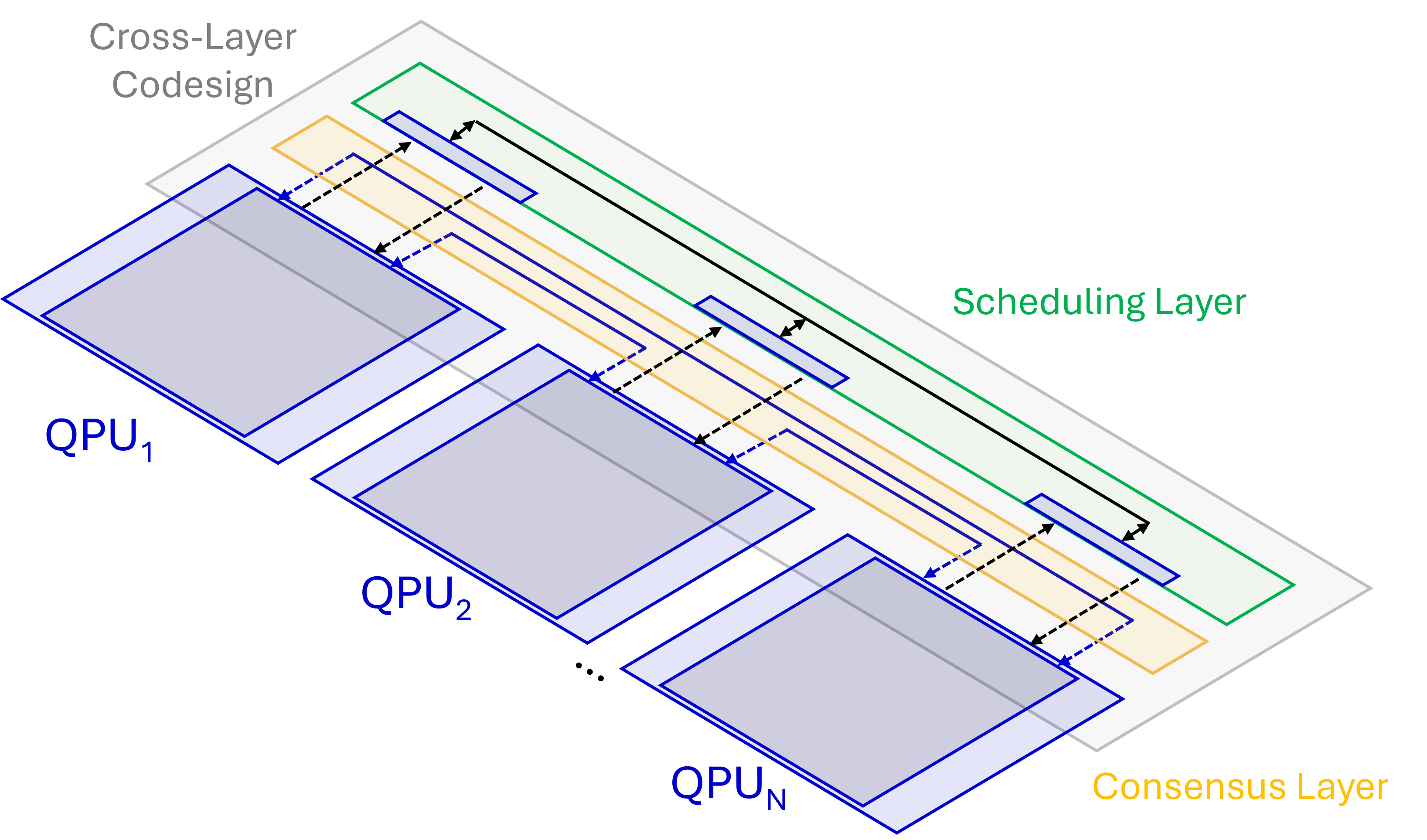}
  \caption{Layered CEAS Control Plane.
Illustration of the CEAS architecture integrating a consensus coordination layer (yellow), an entanglement scheduling layer (green), and a cross-layer signalling interface (grey). Distributed QPUs exchange quantum gradient payloads and classical control metadata over authenticated channels to support synchronised model updates under heterogeneous fidelity constraints.}
  \label{fig:quantum_network_architecture}
\end{figure}

\section{Result}

We emulate a mid-scale distributed quantum--classical learning system comprising $N=50$ compute nodes interconnected via entanglement-enabled links. The network is partitioned into two operational regimes: 60\% of the nodes maintain high-fidelity operation (single- and two-qubit error rates of approximately 1\%) and train on locally separable data characterised by a class separation $\delta_c=1.4$ and label noise $\eta_y=3\%$. In contrast, the remaining 40\% operate in a Byzantine regime, exhibiting elevated error rates (approximately 20\%) and adversarial data distributions ($\delta_c=0.9$, $\eta_y=25\%$). Time advances through 300 discrete optimisation rounds; in each round, every active node refines a local quantum-neural model and subsequently attempts global aggregation. The CEAS protocol combines (i) fidelity-weighted gossip aggregation with logarithmic mixing depth, (ii) a Byzantine-fault-tolerant checkpoint every fourth round governed by Kullback--Leibler divergence thresholds, and (iii) an entanglement scheduler that allocates up to 250 Bell pairs per round following an earliest-expiring-first policy with coherence times $\tau_c \in [3,6]$ rounds.

Each node experiences compound noise processes: quantum decoherence evolves as $\mathcal{F}_{t+1} = \mathcal{F}_{t}\bigl(1-\gamma(\tau_c)\bigr)$, with a time-dependent decay factor $\gamma(\tau_c)$, while Byzantine agents inject randomised perturbations including Gaussian noise $\mathcal{N}\bigl(0,0.6\sigma_W\bigr)$, weight reversal ($W \gets -1.05\,W$), or scaling attacks ($W \gets 1.6\,W$) that adapt dynamically over time. Nodes are quarantined when their rolling divergence $\sum_{k=t-5}^{t}\mathcal{D}_{\mathrm{KL}}^{(k)}$ exceeds a fidelity-dependent threshold $\Theta\bigl(\mathcal{F}_{t}\bigr)$ and may be re-admitted once the active-node count falls below the minimum threshold ($N_{\text{active}}<15$) or their fidelity recovers. For comparison, we perform a baseline experiment in which aggregation nodes are selected uniformly at random, with no fidelity weighting, quarantine enforcement, or entanglement-aware scheduling. Evaluation metrics include global test-set accuracy under coherent conditions ($\delta_c=1.6$, $\eta_y=3\%$), Bell-pair utilisation efficiency, and the cumulative isolation rate of malicious nodes, jointly quantifying resilience against compounded decoherence and Byzantine fault scenarios.

\begin{figure}[!t]
  \centering
  \includegraphics[width=0.5\textwidth]{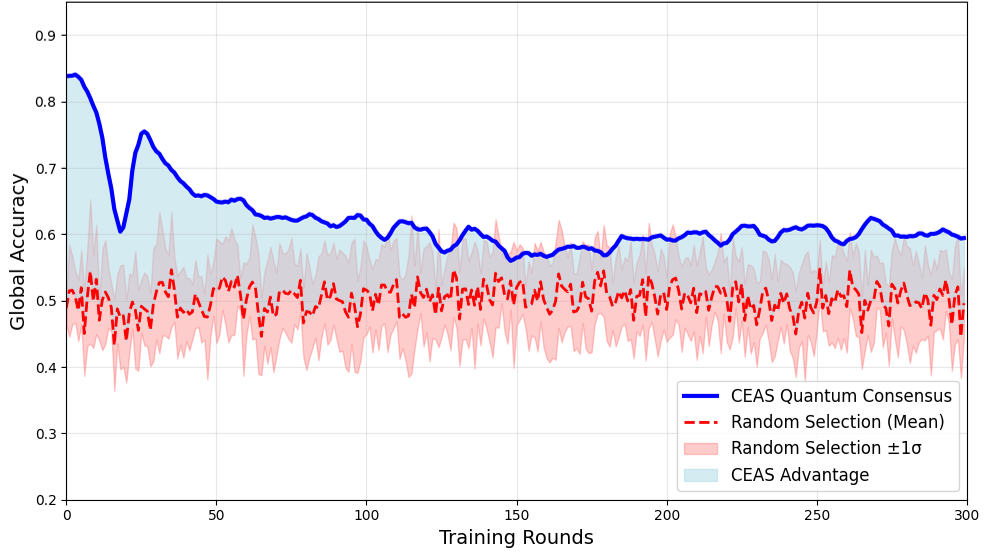}
  \caption{Comparison of global accuracy across training rounds for CEAS quantum consensus versus random participant selection, highlighting the stability and accuracy gains from entanglement-aware scheduling.}
  \label{fig:CEAS_result}
\end{figure}

Fig.~\ref{fig:CEAS_result} tracks global test-set accuracy across all training rounds, highlighting characteristic Byzantine-resilient behaviour. The CEAS protocol (solid blue curve) initially achieves high accuracy (approximately 85\%) based on untampered local models trained on high-quality data from honest nodes. As coordinated Byzantine attacks propagate during the vulnerability window between BFT checkpoints, performance declines to around 60\%. The subsequent recovery to approximately 75\% demonstrates the effectiveness of fidelity-weighted quarantine, which successfully isolates over 90\% of malicious nodes. Accuracy then stabilises between 60\% and 62\%, maintaining a 10--15 percentage point lead over the random-selection baseline (dashed red curve), even as reactivation of previously quarantined nodes and cumulative quantum decoherence ($\Delta\mathcal{F} = \mathcal{F}_0 \prod (1 - \gamma_t)$) exert pressure on convergence. The shaded envelopes confirm that CEAS achieves approximately 3$\times$ lower variance ($\sigma_{\mathrm{CEAS}}\approx0.02$) relative to the baseline, effectively suppressing decoherence noise through cross-layer coordination. Concurrently, sustained Bell-pair utilisation exceeding 90\% within prescribed budgets evidences Pareto-efficient resource allocation that isolates adversarial behaviour without depriving honest participants. Collectively, these dynamics substantiate that entanglement-aware consensus delivers superior accuracy, resilience, and stability within NISQ constraints.

\section{Conclusion}

Consensus-driven, entanglement-aware scheduling lays out a credible path for elevating distributed quantum neural networks from laboratory prototypes to production-grade learning infrastructures. By (i) embedding fidelity weights into the aggregation rule, (ii) synchronising entanglement generation with workload urgency, and (iii) hardening every layer against Byzantine interference, the CEAS framework fuses quantum networking, machine-learning optimisation, and distributed-systems engineering into a single, experimentally-verifiable stack. Turning this blueprint into practice now hinges on three community-level efforts: (a) open benchmarks that pinpoint the operating regime in which quantum consensus overtakes classical BFT; (b) a hardware–software co-roadmap that co-optimises memory lifetimes, entanglement throughput, and protocol time-outs to remove synchronisation bottlenecks; and (c) a security-first deployment model rooted in quantum-secure signatures and continuous syndrome auditing. With these pillars in place, CEAS can deliver the rigorous consensus guarantees that entanglement-rich networks require, supporting reliable, high-throughput training of distributed quantum neural networks and paving the way toward a resilient quantum-Internet ecosystem.

\section*{Acknowledgment}
This work was supported by the Engineering and Physical Sciences Research Council (EPSRC) under grant number EP/W032643/1 and Imperial Quantum ICoNYCh Seed Fund.

\bibliographystyle{ieeetr}
\bibliography{references}

\end{document}